\pdfoutput=1
\documentclass[11pt]{article}
\usepackage[]{acl}

\usepackage{times}
\usepackage{latexsym}
\usepackage{amsmath}
\usepackage{graphicx}
\usepackage{natbib} 
\usepackage{textcomp}
\usepackage{xcolor}
\usepackage{amsmath,amsthm,array}
\usepackage{caption}
\usepackage{subcaption}
\usepackage{algorithm}
\usepackage{algorithmic}
\usepackage{multirow} 
\usepackage{hyperref}
\usepackage{booktabs}

\usepackage[T1]{fontenc}
\usepackage[utf8]{inputenc}

\usepackage{microtype}

\usepackage{inconsolata}

\newcommand{\C}{\mathcal{C}}

\newcommand{\sat}{\mathcal{S}}
\newcommand{\dsat}{\mathcal{D}}

\newcommand{\jn}[1]{}
\renewcommand{\jn}[1]{{\color{red} JN: {#1}}}

\newcommand{\jean}[1]{}
\renewcommand{\jean}[1]{{\color{blue} JL: {#1}}}

\newcommand{\jay}[1]{}
\renewcommand{\jay}[1]{{\color{green} JS: {#1}}}

\newcolumntype{C}[1]{>{\centering\arraybackslash}m{#1}}

\renewcommand{\footnotemark}{}

\newcommand{\modelname}{{\em SPUR}}
\newcommand{\modelfullname}{Supervised Prompting for User satisfaction Rubrics}

\title{Interpretable User Satisfaction Estimation for \\Conversational Systems with Large Language Models}

\author{Ying-Chun Lin$^{*\ddag}$, Jennifer Neville$^{*\dag}$, Jack W. Stokes$^{*\dag}$, Longqi Yang$^{*\dag}$,\thanks{\hspace{-6mm}$^{*}$These corresponding authors contributed equally to this work. Email: lin915@purdue.edu, jenneville@microsoft.com, jstokes@microsoft.com, longqi.yang@microsoft.com} \\ {\bf Tara Safavi}$^{\dag}$,
{\bf Mengting Wan}$^{\dag}$, {\bf Scott Counts}$^{\dag}$, {\bf Siddharth Suri}$^{\dag}$, \\ {\bf Reid Andersen}$^{\dag}$,
{\bf Xiaofeng Xu}$^{\dag}$, {\bf Deepak Gupta}$^{\dag}$, {\bf Sujay Kumar Jauhar}$^{\dag}$,\\  {\bf Xia Song}$^{\dag}$, {\bf Georg Buscher}$^{\dag}$, {\bf Saurabh Tiwary}$^{\dag}$, {\bf Brent Hecht}$^{\dag}$, {\bf Jaime Teevan}$^{\dag}$\\
        $^{\dag}$Microsoft Corporation, $^{\ddag}$Purdue University\\
        }

\begin{document}
\maketitle

\begin{abstract}
Accurate and interpretable user satisfaction estimation (USE) is critical for understanding, evaluating, and continuously improving conversational systems. Users express their satisfaction or dissatisfaction with diverse conversational patterns in both general-purpose (ChatGPT and Bing Copilot) and task-oriented (customer service chatbot) conversational systems. Existing approaches based on featurized ML models or text embeddings fall short in extracting generalizable patterns and are hard to interpret. In this work, we show that LLMs can extract interpretable signals of user satisfaction from their natural language utterances more effectively than embedding-based approaches. Moreover, an LLM can be tailored for USE via an iterative prompting framework using supervision from labeled examples. Our proposed method, \modelfullname{} (\modelname{}), not only has higher accuracy but is more interpretable as it scores user satisfaction via learned rubrics with a detailed breakdown.
\end{abstract}

\section{Introduction}

General-purpose conversational systems such as ChatGPT and Copilot are revolutionizing how people live and work. Understanding when and why users are satisfied or dissatisfied is critical for the continuous improvement of these systems. It helps system developers identify areas of improvements, conduct effective A/B experiments, and optimize underlying models. Unsurprisingly, developing machine learning models for User Satisfaction Estimation (USE)~\citep{urgo, use_cl, sentiment, rq_use, rq_use2} has captured significant attention from the research community.

\begin{figure}[t]
  \centering
  \includegraphics[width=\linewidth]{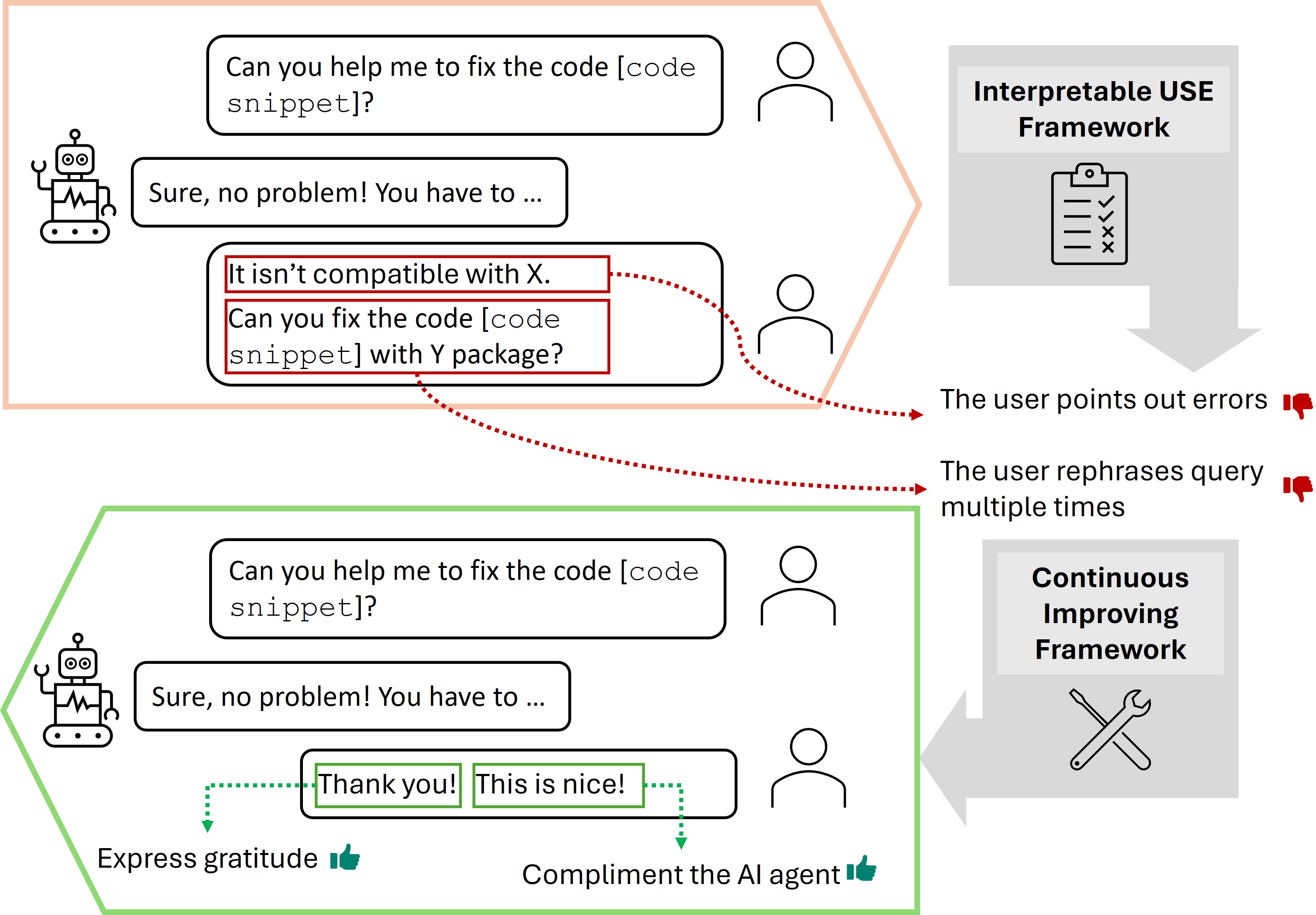}
  \vspace{-2mm}  
  \caption{Illustration of user utterances with satisfaction patterns (green) and dissatisfaction patterns (red).}
  \vspace{-2mm}
\label{fig:example}
\end{figure}

When estimating user satisfaction, simply classifying that a user is satisfied or dissatisfied is insufficient. Understanding the reason why a user is satisfied or dissatisfied is just as valuable. For example, frequent query reformulation presents opportunities for prompt recommendation and conversations where users explicitly correct a bot's mistakes can suggest examples for model alignment. See Figure~\ref{fig:example} for an illustration. However, most existing work has focused on improving  classification accuracy and has overlooked  interpretability. Representation learning-based approaches~\citep{use_speaker, usda, asap} are relatively opaque due to their use of neural models (e.g., embeddings) and thus offer little insight into conversational patterns that indicate satisfaction/dissatisfaction. Similar limitations apply to reward models for training LLMs, e.g., RLHF~\citep{rlhf} and RLAIF~\citep{rlaif}. In this case, the learned model produces a continuous ``reward'' score that aims to distinguish outputs that a human prefers without explaining why a conversation has a higher score than others. To our knowledge, these reward models have not been directly used for USE, but we treat it as a baseline due to their ability to rank outputs with respect to human preferences.

Some prior work addressed the interpretation needs of USE via featurized ML models. Examples include \citet{paradise}, which evaluated user satisfaction based on human-annotated features assessing task success and dialogue costs, and \citet{rq_use}, which proposed domain-independent features that evaluate response quality. However, the growth of LLM-based conversational systems (e.g., ChatGPT, Bing Copilot) means user queries in conversational systems may now be across multiple domains and intents (e.g., task-oriented, QA, chitchat). As such, approaches based on domain-specific features have limited generalizability to these diverse conversational patterns~\citep{survey_dialogue_sys}.

In this work, we make the key observation that LLMs can achieve both high classification accuracy and fine-grained interpretability at the same time -- through their ability to reason about user conversational patterns and identify salient pattern classes that generalize and produce accurate predictions. We propose \modelfullname{} (\modelname). We consider a {\em few-shot} scenario, where a small number of training examples are available, and develop a supervised, iterative prompting framework that uses an LLM to (1) extract signals of  satisfaction from user utterances in a labeled training set, (2) summarize the reasons into rubrics for identifying satisfaction/dissatisfaction conversational patterns, and (3) apply the rubrics to predict satisfaction labels on unseen conversations.

In addition to being more accurate, our approach provides an interpretable rubric for understanding the conversational patterns that indicate user satisfaction/dissatisfaction. Notably, our approach can be used to learn SAT/DSAT patterns automatically for different conversational systems. In our experimental results, we show the distributions of patterns in different types of systems and demonstrate how these patterns (1) correlate to overall user satisfaction, and (2) differ across domains. 

Moreover, we show that we can \textit{scale} the application of the learned rubrics in two ways. First, we show that we can {\em distill} individual rubric items into an embedding-based model that can be applied at scale without the need for LLM prompting. Next, we show that we can add rubric items as features to an embedding-based model to increase the accuracy of embedding-only models on datasets with more available training data.

The main contributions of our work include:
\begin{itemize}
    \item We propose \modelfullname{} (\modelname{}), a novel framework for estimating user satisfaction in conversational systems with LLMs.
    \item We show the \modelname{} prompting process extracts patterns into clear and interpretable rubrics that guide the LLM to classify user satisfaction and show   that diverse rubrics are learned automatically for different domains.
    \item We show \modelname{} outperforms existing methods across different types of conversational systems when training data is limited and provide insights into the factors that influence user satisfaction.
    \item We use knowledge distillation to scale the application of  learned rubrics and show the rubrics can continuously improve performance %on USE 
    as more training data is available.
\end{itemize}

\section{Problem Definition and Related Work}
\noindent\textbf{Problem Definition.} Let a conversation $C$ from session $i$ and consisting of $t$ interaction turns of user-agent utterances be $C_i = [ U_1, A_1, \hdots, U_t, A_t ]$. Here $U_t$ refers to a {\em user} utterance and $A_t$ refers to an {\em AI agent} utterance. The user-agent utterances $C_i$ typically consist of multiple turns, e.g., $t>1$. The conversation also has an overall user satisfaction label $y_i \in [-1,+1]$ provided by {\em thumb feedback} (e.g., like or dislike). 

Our goal is to learn a function $f : C \rightarrow y$ to accurately predict the satisfaction label of unseen conversations and explain the predicted label. In multi-turn conversational sessions, a user can convey their satisfaction (or dissatisfaction) explicitly in their utterances or implicitly through their behavioral interactions with the agent. We refer to these  satisfaction/dissatisfaction {\em conversational patterns} as SAT/DSAT patterns. Let $\sat=\{s_1, s_2, \cdots, s_{\infty}\}$ and $\dsat=\{d_1, d_2, \cdots, d_{\infty}\}$ be the set of all interpretable SAT and DSAT patterns respectively. We assume these are latent and unknown. The goal is to identify a subset of SAT/DSAT pattern classes ($\sat_s \subset \sat,\dsat_s \subset \dsat$) that summarize the conversation enough to accurately predict its label: 
$P(y | C) \approx P\Big(y \: \Big| \: \sat_s(C), \dsat_s(C) \Big)$.

SAT and DSAT patterns may be direct compliments or complaints about the AI agent's responses, or behavioral patterns that implicitly express user satisfaction. For example, users may continue to ask follow-up questions, indicating that the AI has provided accurate information that inspires their curiosity and leaves them satisfied. Conversely, if a user repeatedly rephrases the same question, it can signal dissatisfaction.

\vspace{2mm}
\noindent\textbf{Related Work.} Numerous prior research studies have examined User Satisfaction Evaluation (USE) through the lenses of sentiment analysis~\citep{use_speaker, sentiment}, content analysis~\citep{paradise, uss}, and response quality assessment~\citep{iq, rq_use}. 
While analyzing user sentiment distribution in a dialogue session can enhance the model's USE capabilities, it is important to note that sentiment analysis is not equivalent to USE~\citep{use_speaker}. Another common approach involves content analysis, which typically necessitates the employment of human annotators to evaluate interaction quality in a dialogue session~\citep{iq, rq_use}. Afterwards, a classifier is trained to predict user satisfaction based on the features extracted from the annotation process.

With the advancement of language models, there is a growing trend in the use of text embeddings to estimate user satisfaction for conversational systems~\citep{use_1, use_2, use_3, uss}. This approach is also being employed to simulate user satisfaction. Some work has focused on identifying dialogue acts or user intents in measuring the fulfillment of the user’s goals~\citep{use_rec, uss}. Other work has focused on incorporating the sequential dynamics of dialogue acts~\citep{usda}, jointly predicting sentiment and satisfaction~\citep{use_speaker}, or modeling  dynamics of satisfaction across turns~\citep{asap}.

Recently, Large Language Models (LLMs) revolutionized the traditional learning framework~\citep{zero_shot_reasoner, cot}, especially for natural language processing (NLP) tasks. LLMs have achieved performance comparable to supervised baselines or state-of-the-art results across various NLP tasks with In-Context Learning (ICL). By providing a few examples or hints~\citep{fewshot,carp} and simple reasoning process~\citep{zero_shot_reasoner, cot}, LLMs can provide significant performance boosts for NLP tasks. \citet{urgo} further uses LLMs as a user simulator for USE and adopts the user simulator into RLAIF~\citep{rlaif} for fine-tuning the existing LLM models. For USE with zero-shot prompting~\citep{zero_shot_reasoner, urgo}, instructions provided by a human may not fit the actual conversation patterns in the data and hence introduce bias. For few-shot prompting~\citet{fewshot,carp}, the provided examples are not enough to describe the full distribution of the conversational patterns, and this results in inaccuracies for USE.

\section{\modelname}
\vspace{-3mm}

\begin{figure*}[ht]
  \centering
  \includegraphics[width=6in]{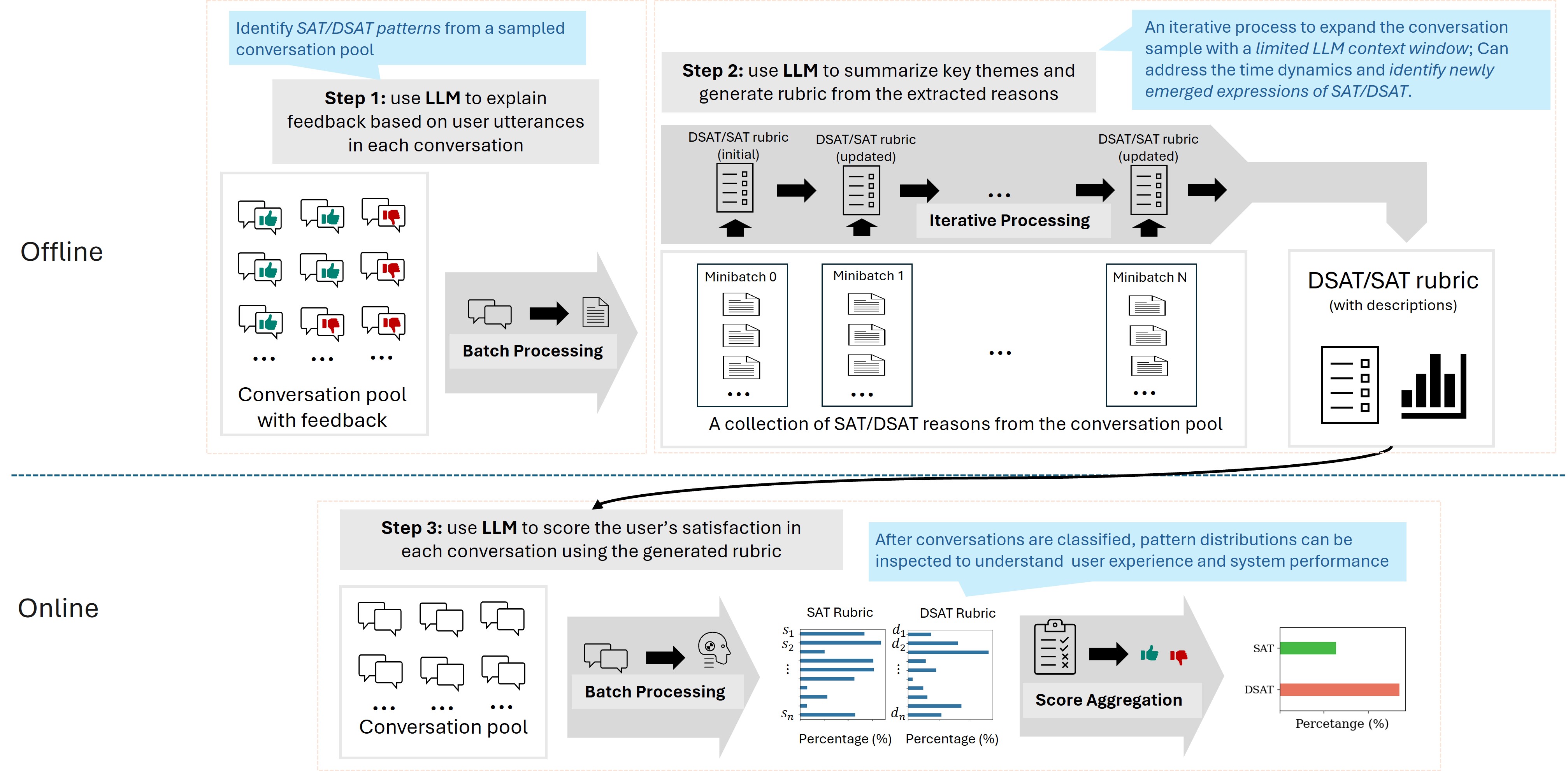}
  \vspace{-2mm}  
  \caption{Illustration of \modelname{} approach. Step 1 corresponds to Sec.~\ref{sec:se}, Step 2: Sec.~\ref{sec:rs}, and Step 3: Sec.~\ref{sec:sat_score}.}
\label{fig:approach}
\vspace{-2mm}
\end{figure*}

We propose {\modelname} for interpretable User Satisfaction Estimation (USE) given a small set of labeled conversation $C$ from a conversational system. Due to the multi-turn and general-purpose nature in such conversational systems, users demonstrate a variety of response patterns when expressing satisfaction or dissatisfaction. Our approach follows the three-phase prompting strategy  depicted in Figure \ref{fig:approach}: Supervised Extraction, Rubric Summarization,  and User Satisfaction Estimation. 
Our three-phase approach is essential for ensuring accuracy, generalization, and interpretability. Through Supervised Extraction, {\modelname} improves accuracy by capturing the diverse conversational patterns in the training set $\C_{train}=\{C_1, C_2, \cdots, C_N\}$, which are annotated with thumb feedback. In the Rubric Summarization stage, the LLM improves generalization and interpretability by identifying prominent SAT and DSAT pattern classes among the full set of extracted pattern.  Finally, {\modelname} uses the learned rubrics generated from the previous stage to score user satisfaction on unlabeled conversations.
For the ease of understanding, we use mathematical definitions to approximate the process of \modelname{} in the following three sections.

\subsection{Supervised Extraction}
\label{sec:se}

The first step of our framework is {\em Supervised Extraction}---where we use a prompt to obtain meaningful and interpretable SAT/DSAT pattern classes from GPT-4, which has an exceptional ability for natural language understanding and reasoning~\citep{opt_explanation, faith_explanation, zero_shot_reasoner}. 
Given a conversation $C_i$ with its user satisfaction label $y_i=+1$, how the user expresses satisfaction in $C_i$  can be formulated as:
$\widehat{\mathbf{s}}_i \approx \underset{s \in \sat}{\operatorname{arg\, max_k}} \; P(\sat|C_i, y_i=+1)$
where $\sat=\{s_1, s_2, \cdots, s_{\infty}\}$ is the set of all possible SAT patterns. The goal is to identify the top-k potential pattern classes $\widehat{\mathbf{s}}_i=\{s_1, s_2,\cdots, s_k\} \subset \sat$ that are exhibited in $C_i$ relevant to satisfaction expression. Similarly,
$\widehat{\mathbf{d}}_i \approx \underset{d \in \dsat}{\operatorname{arg\, max_k}} \; P(\dsat|C_i, y_i=-1)$,
where $\dsat=\{d_1, d_2, \cdots, d_{\infty}\}$ is the set of all possible DSAT patterns. 

The prompt for generating the possible $\widehat{\mathbf{s}}$ or $\widehat{\mathbf{d}}$ patterns from $\C_{train}$ is provided in Appendix~\ref{apx:se_prompt}. In our prompt, we specifically require GPT-4 to restrict $k \leq 3$ for each $C_i$. The prompt for DSAT patterns is similar; we only replace ``satisfaction'' with ``dissatisfaction'' in the instructions.

For the ease of discussion in the next section, let $\widehat{\sat} = \{\widehat{s}_1,\cdots,\widehat{s}_N\}$ denote all the SAT patterns derived from Supervised Extraction and $\widehat{\dsat} = \{\widehat{d}_1,\cdots,\widehat{d}_N\}$ are all the  DSAT patterns.

\subsection{Rubric Summarization}
\label{sec:rs}

The patterns extracted through Supervised Extraction prompting may exhibit significant variation based on the text descriptions across different conversations, and their relative importance may not be uniform. Our observations indicate that, despite differences in the text descriptions, most $\widehat{s}_i\in\widehat{\sat}$ and $\widehat{d}_i\in\widehat{\dsat}$ are semantically similar. As such, the goal of the Rubric Summarization stage is to further condense $\widehat{\sat}$ and $\widehat{\dsat}$, and identify frequently occurring SAT/DSAT patterns across $\C_{train}$. The outcome of this process is the establishment of a clear rubric for USE based on $\widehat{\sat}$ and $\widehat{\dsat}$.

However, it is infeasible to summarize $\widehat{\sat}$ and $\widehat{\dsat}$ into a clear rubric using a single prompt because the number of tokens in $\widehat{\sat}$ and $\widehat{\dsat}$ is too large to fit into the context size limit of GPT-4. (Note, we used GPT-4-32K with a 32K context window in this work.) To address this, we propose an iterative process to incrementally update the satisfaction and dissatisfaction rubrics by processing a fixed-size minibatch of patterns. The satisfaction batches are denoted as $\{\widehat{\sat}_1, \widehat{\sat}_2, \cdots, \widehat{\sat}_B\}$ where $\widehat{\sat}=\cup_{b=1}^{B}\widehat{\sat}_b$ and the number of batches is $B$. Similarly, $\{\widehat{\dsat}_1, \widehat{\dsat}_2, \cdots, \widehat{\dsat}_B\}$ are the batches to learn the dissatisfaction rubric and $\widehat{\dsat}=\cup_{b=1}^{B}\widehat{\dsat}_b$. In each iteration, GPT-4 is asked to generate an $n$-item rubric for the SAT patterns in $\widehat{\sat}_b$. This $n$-item SAT rubric is then appended at the end of $\widehat{\sat}_{b+1}$ to incorporate in the generation of the next $n$-item SAT rubric. The iterative process continues until the final batch, and then the last output $n$-item rubric is used as the final SAT rubric $\widetilde{\sat}=\{\tilde{s}_1\cdots\tilde{s}_{n}\}$. The process is illustrated at Step 2 in Figure~\ref{fig:approach}. A similar process is applied to generate the DSAT rubric $\widetilde{\dsat}=\{\tilde{d}_1\cdots\tilde{d}_{n}\}$. We set $n=10$ in our experiments. The final SAT and DSAT rubrics for Bing Copilot are in Table~\ref{tab:copilot_rubric}, and the Rubric Summarization prompt is provided in Appendix~\ref{apx:sr_prompt}.

There are two benefits to utilizing the LLM-generated satisfaction and dissatisfaction rubrics from this iterative process. First, the rubrics are developed in a supervised manner from the set of training conversations, $\C_{train}$, thereby ensuring that prominent (and thus predictive) SAT and DSAT pattern classes in the distribution are identified. As a result, the generated rubrics provide a clear guideline for GPT-4 to estimate user satisfaction accurately. Second, the rubrics are generated from more examples than can fit in a single context window. As such, Rubric Summarization improves the generalization for GPT-4 in terms of in-context learning.

\subsection{User Satisfaction Estimation}
\label{sec:sat_score}

After learning the satisfaction rubric $\widetilde{\sat}$ and dissatisfaction rubric $\widetilde{\dsat}$, we incorporate the generated rubrics as instructions in a third prompt that we provide GPT-4 to score user satisfaction. The rubric items provide a consistent decision making criteria and enhance the performance of GPT-4 on USE. For each rubric item $\tilde{s}_r\in\widetilde{\sat}$ or $\tilde{d}_r\in\widetilde{\dsat}$, the prompt asks GPT-4 to make a binary decision as to whether a given conversation demonstrates the described behavior. If the answer is "Yes", the prompts further instruct GPT-4 to evaluate how likely the expressed pattern will impact the user's overall satisfaction/dissatisfaction with their interaction on a scale of $1-10$ (low to high). Otherwise, if the answer is ``No,'' the score is $0$. After  the score for each rubric item is output, we further aggregate the scores into a single SAT score $\mathcal{R}$ to represent the overall user satisfaction in the given conversation. $\mathcal{R}$ is computed as:
$\mathcal{R} = \sum_{i = 1}^{n} \tilde{r}_{s_i} - \sum_{j = 1}^{n} \tilde{r}_{d_j}$
where $\tilde{r}_{s_i}$ is the score for the $i$th SAT rubric item and $\tilde{r}_{d_j}$ for the $j$th DSAT item. The prompt is in Appendix~\ref{apx:use_prompt}.

\section{Evaluation}

We evaluate \modelname{} by comparing its performance quantitatively against previous embedding-based approaches and several ablated versions of our LLM-based approach. 

\subsection{Baselines}
We compare \modelname{} with two LLM-based methods, including ZeroShot and FewShot, and three embedding-based methods: Linear Regression, USDA~\citep{usda} and ASAP~\citep{asap}. Note that we choose GPT-4 for all LLM-based methods instead of other smaller language models because smaller language models struggle to accurately generate scores for each rubric item, which results in incorrect SAT scores. The detailed descriptions of the models are as follows:
\begin{enumerate}
    \item Lin-ada: Linear regression model with ada-002 embedding~\cite{openai_embeddings}
    \item USDA~\citep{usda}\footnote{\url{https://github.com/dengyang17/USDA}} is an embedding-based method for USE by jointly optimizing user satisfaction and the sequential dynamics of dialogue acts.
    \item ASAP~\citep{asap}\footnote{\url{https://github.com/smartyfh/ASAP}} is another embedding-based method which models user satisfaction across turns via a Hawkes Process.
    \item Zero shot: prompt GPT-4 directly to score conversations for user satisfaction with basic reasoning steps by providing explanations.
    \item Few shot: prompt GPT-4 directly to score conversations, include 2 examples of labeled conversations to guide GPT-4 to determine user satisfaction and include basic reasoning steps by providing explanations.
    \item RQ: prompt GPT-4 with a manually selected features to assess the response quality~\citep{rq_use2} and ask GPT-4 to determine user satisfaction based on the set of features.
    \item Reward: pretrained reward model for RLHF\footnote{https://huggingface.co/OpenAssistant/reward-model-deberta-v3-large-v2}.
\end{enumerate}

\subsection{Dataset}
We use four datasets to evaluate the performance of the compared methods. Bing Copilot is a general-purpose and multilingual conversational system, and this dataset includes 50K fully anonymized conversations.\footnote{All personal, private, or sensitive information was scrubbed and masked before the conversations were used for this research. The access to the dataset is strictly limited to the authors who conducted hands-on analysis and model development.}. MWOZ~\citep{mwoz}, SGD~\citep{sgd} and ReDial~\citep{redial} are three task-oriented, English conversational systems, and they have 1155, 1638, and 1387 conversations, respectively. These three datasets are further processed and labeled user satisfaction by \citet{uss}. Because \citet{uss} labeled user satisfaction by turn, we further process these labels into a label to represent the overall satisfaction of the whole conversation. The preprocessing details are described in Appendix~\ref{apx:preprocess}.

\vspace{2mm}
\noindent\textbf{Ethics.} As part of the production process, the Bing Copilot data is anonymized, and each conversation is formed by aggregating turns based on a unique conversation ID. Thus, none of the researchers who analyzed the data are able to recover and identify the conversations from any individual user. In addition, this research study was reviewed and approved by representatives from our institutional review board (IRB), as well as our ethics and security teams. No formal IRB certificate was required since we did not conduct any human studies for this paper.

\begin{table*}[htbp]
\caption{Precision (P), Recall (R), and F1 Score (F1) on USE with small training set sizes. The training sizes are shown besides the name of each dataset. The testing size is $80\%$ of the data. The best scores are in \textbf{bold face}.}
\begin{footnotesize}
\begin{center}

\begin{tabular}{p{4em}p{1.6em}p{1.6em}p{1.6em}p{1.6em}p{1.6em}p{1.6em}p{1.6em}p{1.6em}p{1.6em}p{1.6em}p{1.6em}p{1.6em}}
\hline
\multirow{2}{*}{Models} & \multicolumn{3}{c}{Bing Copilot ($0.8\%$)} & \multicolumn{3}{c}{MWOZ ($5\%$)} & \multicolumn{3}{c}{SGD ($5\%$)}  & \multicolumn{3}{c}{ReDial ($5\%$)}\\ \cmidrule(l{0.6em}r{0.6em}){2-4}\cmidrule(l{0.6em}r{0.6em}){5-7}\cmidrule(l{0.6em}r{0.6em}){8-10}\cmidrule(l{0.6em}r{0.6em}){11-13}
 & P & R & F1 & P & R & F1 & P & R & F1 & P & R & F1\\

\hline
Lin-ada & 74.0 & 72.5 & 73.3 & 48.0 & 24.0 & 29.1 & 53.9 & 34.9 & 39.4 & 56.0 & 27.7 & 33.6 \\
USDA & 49.7 & 53.3 & 47.3 & 38.1 & 50.7 & 35.3 & 66.1 & 66.3 & 61.3 & 56.1 & 56.8 & 48.3 \\
ASAP & 66.0 & 70.1 & 58.4 & 51.2 & 56.1 & 52.5 & 64.8 & 69.8 & 66.3 & 60.0 & 63.6 & 58.3 \\
\hline
Reward & 43.6 & 52.0 & 52.7 & 63.0 & 47.4 & 40.7 & 65.3 & 66.9 & 58.6 & 44.1 & 57.7 & 48.2 \\
ZeroShot & 75.5 & 73.7 & 68.3 & 66.2 & 52.2 & 53.6 & \textbf{75.3} & 70.8 & 71.9 & \textbf{71.2} & 58.2 & 57.0 \\
FewShot  & 63.8 & 68.8 & 61.9  & \textbf{68.4} & 47.4 & 44.8 & 67.3 & 69.9 & 66.2 & 41.0 & 62.2 & 49.4  \\
RQ & 57.7 & 69.6 & 57.7 & 33.3 & 52.5 & 38.1 & 49.6 & 67.3 & 54.4 & 40.6 & 63.6 & 49.5 \\
\hline
% SESRP (Label) & 44.6 & 69.7 & 44.6 & 53.6 & 54.5 & 80.5 & 54.5 & 62.5 & 50.2 & 78.0 & 50.2 & \textbf{50.5} & 43.6 & 73.0 & 43.6 & 52.8 \\
\modelname{} & \textbf{76.3} & \textbf{77.2} & \textbf{75.4} & 65.7 & \textbf{61.6} & \textbf{59.0} & 73.7 & \textbf{74.1} & \textbf{72.6} & 68.4 & \textbf{68.7} & \textbf{66.3}
 \\
\hline
\end{tabular}
\label{tab:use_fewshot}
\end{center}
\end{footnotesize}

\end{table*}
\vspace{2mm}
\subsection{USE under Few-Shot Setting.}
Table~\ref{tab:use_fewshot} shows the performance of each model trained with a small number of training examples. The performance scores are the average of five runs in different train/test splits. The performance metrics are weighted based on the label distributions due to the data imbalance in the different datasets. The training set sizes are shown beside the name of each dataset, and the remaining $80\%$ of the data is used for testing. The number of items in the satisfaction and dissatisfaction rubrics is ten, respectively. Three task-oriented datasets have larger training sizes because we want to ensure that there are at least ten conversations with satisfaction labels and ten conversations with dissatisfaction labels to derive \modelname{}'s rubrics.

The performance difference between ZeroShot and \modelname{} lies in that the learned rubrics can provide better guidance for LLMs to determine user satisfaction. Comparing the performance between RQ and \modelname{} in Table~\ref{tab:use_fewshot}, the effectiveness of the rubrics can be observed. Prompting with learned rubrics can provide guidance specific to a dataset for LLMs than prompting with a set of manually selected features~\cite{rq_use2} used by all datasets. On the other hand, FewShot has worse performance compared to other methods because the examples provided in the prompt cannot cover many types of satisfaction/dissatisfaction conversational patterns, and the decision is usually biased by the examples provided in the prompt.

The performance of the Reward model~\citep{hugging_reward} validates our hypothesis that Reward models used for RLHF is not a good proxy for scoring user satisfaction. Because Reward models are usually trained with auxiliary human feedback, this reward is not learned from the perspective of the user who was involved in the conversation with the AI agent~\citep{rlhf_risk}.

Embedding methods perform worse than \modelname{} in 
Table~\ref{tab:use_fewshot}. Due to the smaller training size, embedding methods cannot generalize well, particularly when the class is imbalanced. They usually have lower weighted F1 scores because they cannot accurately classify the conversations from the less likely classes. However, the strong performance of ZeroShot and \modelname{} demonstrate that LLM-based methods can effectively identify accurate satisfaction/dissatisfaction conversational patterns  from limited data.

\begin{table}[tbp]
\caption{The F1 Gain shows the improvement after learning the dataset-specific rubrics compared to the Bing Copilot rubrics, and the last two columns report the set difference between the SAT/DSAT rubrics of each open dataset and the Bing Copilot dataset.}
\begin{footnotesize}
\begin{center}
\begin{tabular}{l>{\small}c>{\small}c>{\small}c}
\hline
\multirow{2}{*}{Dataset}  & F1 & Num. New &  Num. New  \\
  & Gain & SAT Patterns & DSAT patterns \\
\hline
MWOZ & $20.8\%$ & 6 & 8 \\
SGD & $9.5\%$ & 3 & 4 \\
ReDial & $9.2\%$ & 5 & 4 \\
\hline
\end{tabular}
\vspace{-3mm}
\label{tab:cross_domain}
\end{center}
\end{footnotesize}
\end{table}

%$\widetilde{\sat}_{(\cdot)}\setminus\widetilde{\sat}_{\text{Copilot}}$
%$\widetilde{\dsat}_{(\cdot)}\setminus\widetilde{\dsat}_{\text{Copilot}}$
\vspace{2mm}
\subsection{Importance of Rubric Summarization.}
Table~\ref{tab:cross_domain} demonstrates that learning the rubric on each dataset is important for improving the performance on USE. In this experiment, we first use the rubric learned from Bing Copilot (Appendix~\ref{apx:use_prompt}) in the prompt for MWOZ, SGD and Redial, and evaluate USE performance. Then, we apply the specialized rubrics learned from the target datasets and reevaluate USE performance to gauge how much the Bing Copilot rubrics fail to generalize across tasks. The weighted F1 scores in the first column show that rubrics learned on domain-specific data produce an average gain of $13\%$. The last two columns show the set difference between the rubric items in the target and source sets, i.e., $\widetilde{\sat}_{(\cdot)}\setminus\widetilde{\sat}_{\text{Copilot}}$ and $\widetilde{\dsat}_{(\cdot)}\setminus\widetilde{\dsat}_{\text{Copilot}}$. Values $\geq 0$ indicate that the Rubric Summarization process learns a different set of SAT/DSAT rubrics compared to that of Bing Copilot. This demonstrates that the handcrafted features used by several previous studies~\citep{paradise, rq_use, rq_use2} are unlikely to generalize across different types of conversational systems. At the same time, manually designing rubrics (features) for each different conversational systems is time consuming and likely to be ineffective. With our LLM Rubric Summarization process, a targeted set of rubric items can be \textit{learned} for each task/domain, thereby improving USE accuracy.

\begin{figure*}[t]
  \centering
  \vspace{-3mm}
  \includegraphics[width=\linewidth]{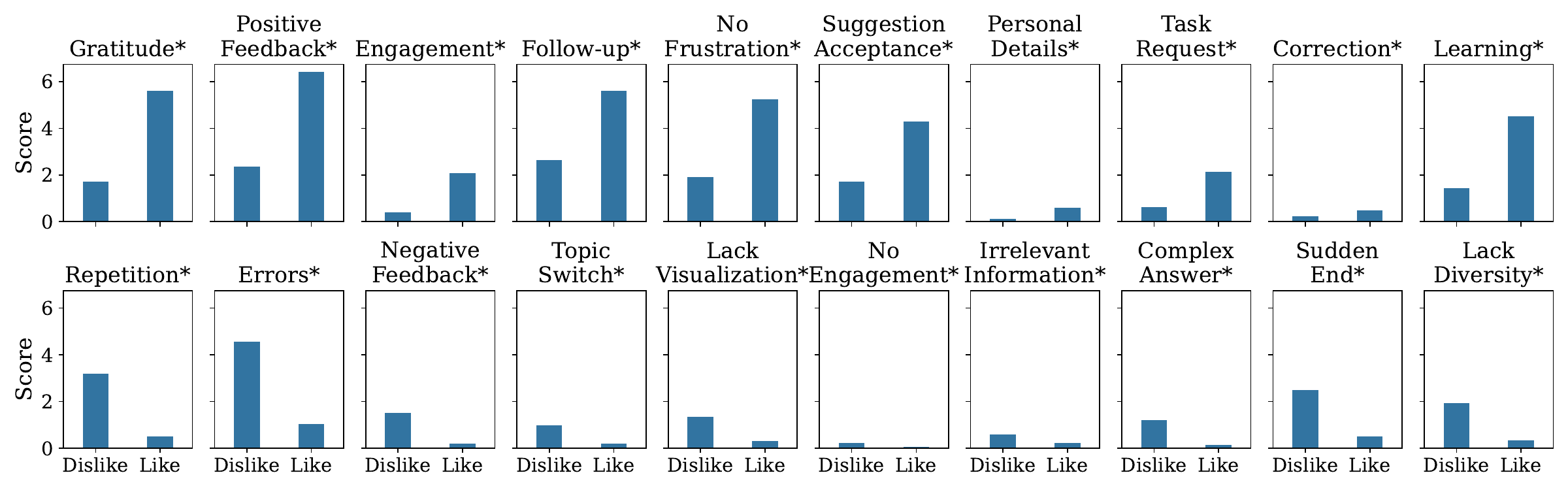}
  \vspace{-6mm}
  \caption{The average scores for each rubric item w.r.t. thumb feedback (Like or Dislike). The `*' beside each keyword indicates that the rubric item is significantly correlated with thumb feedback.}
  \vspace{-1mm}
  \label{fig:rubric_utility}
\end{figure*}
\vspace{2mm}
\subsection{Rubric vs. Thumb Feedback.}
Figure~\ref{fig:rubric_utility} shows the correlation between each rubric item and thumb feedback from users. As discussed in Section~\ref{sec:sat_score}, we ask GPT-4 to generate a label (Yes or No) and a score (0 to 10) for each rubric item in the prompt. The ``Yes'' label for a rubric item means that the conversational pattern exists in the given conversation, and the score indicates how likely this conversation pattern impacts the overall user satisfaction. The title of each sub-figure in Figure~\ref{fig:rubric_utility} provides a short keyword to summarize the rubric item, and the full descriptions of these keywords are listed in Table~\ref{tab:copilot_rubric}. The x-axis shows thumb feedback from users (Like or Dislike). The y-axis shows the average score for each rubric item with respect to the conversations with particular user satisfaction labels. The satisfaction rubric items, which are in the top row, have a higher average score when thumb feedback is Like. Conversely, the conversations where thumb feedback is Dislike have higher scores for the dissatisfaction rubric items (bottom row).

From Figure~\ref{fig:rubric_utility}, we can see that all twenty rubric items exhibit a significant difference in scores with respect to thumb feedback.  This indicates that the score for each rubric item can be used to improve USE predictions. We conducted a Chi-Square test between the labels of each rubric item and thumb feedback from users to observe whether these rubric items are useful for USE. The ``*'' beside each keyword indicates that the rubric item is significantly correlated ($p<0.05$?) with the signals provided by thumb feedback.

\vspace{2mm}
\subsection{Pattern Variance for Different Conversational Systems.}
\begin{figure}[tbh]
\begin{subfigure}{.5\linewidth}
  \centering
  \includegraphics[trim={0 0 0 0},clip,width=\linewidth]{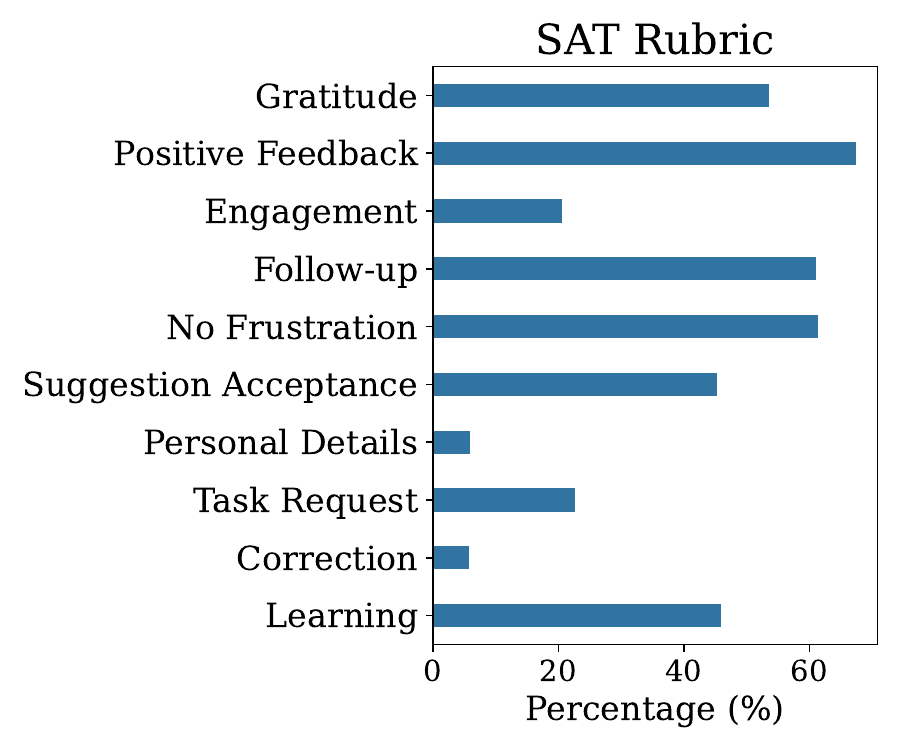}
  \caption{Bing Copilot.}
  \label{fig:bing_sat_feature_dist}
\end{subfigure}%
\begin{subfigure}{.5\linewidth}
  \centering
  \includegraphics[trim={0 0 0 0},clip,width=\linewidth]{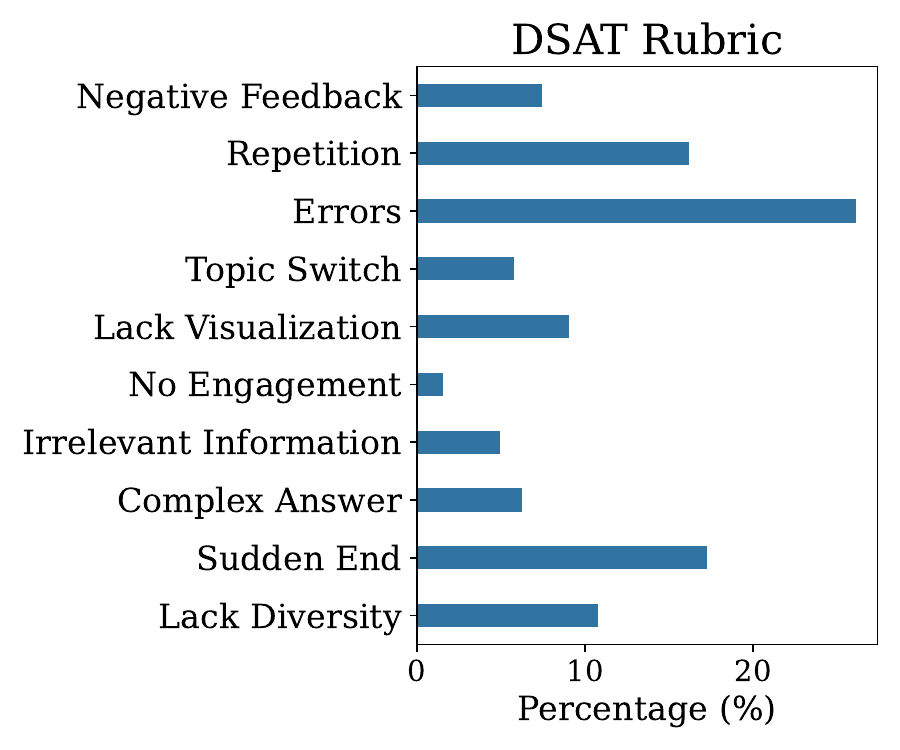}
  \caption{Bing Copilot.}
  \label{fig:bing_dsat_feature_dist}
\end{subfigure}\\
\begin{subfigure}{.5\linewidth}
  \centering
  \includegraphics[trim={0 0 0 0},clip,width=\linewidth]{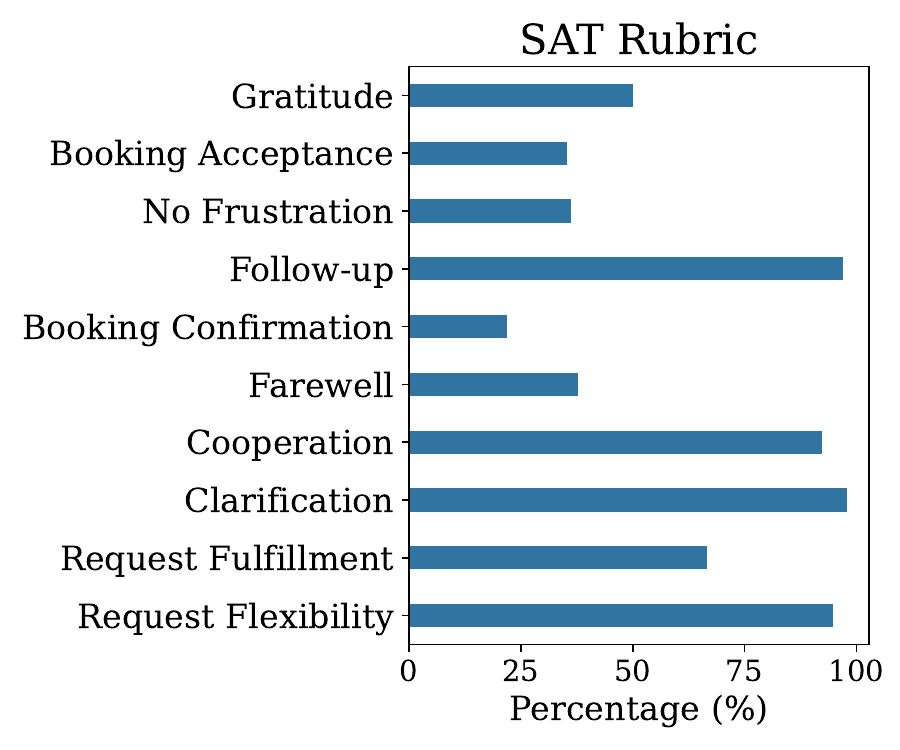}
  \caption{MWOZ.}
  \label{fig:mwoz_sat_feature_dist}
\end{subfigure}%
\begin{subfigure}{.5\linewidth}
  \centering
  \includegraphics[trim={0 0 0 0},clip,width=\linewidth]{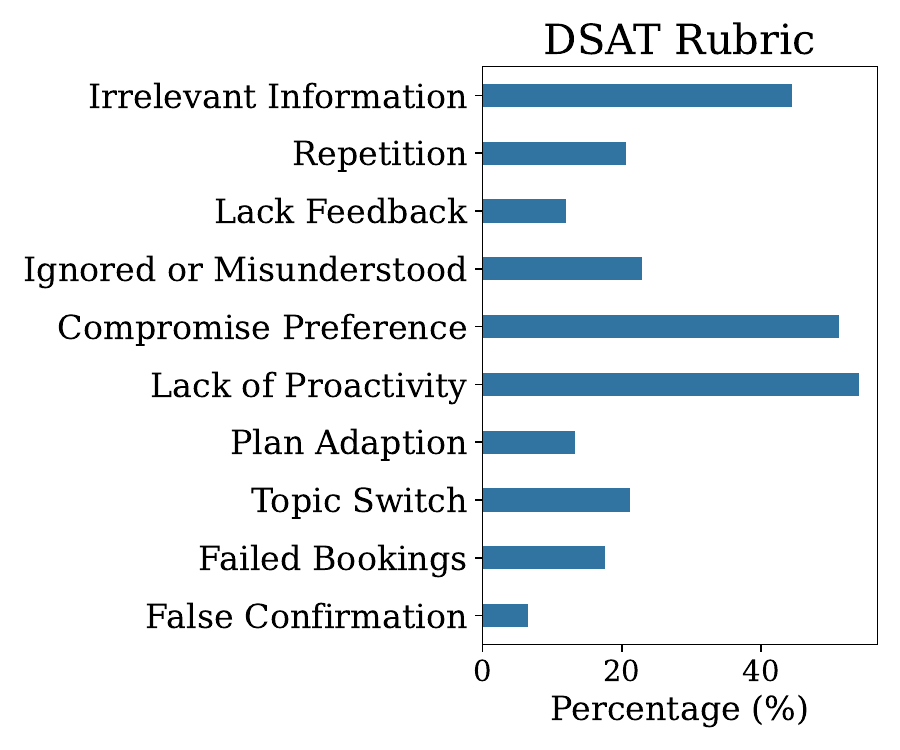}
  \caption{MWOZ.}
  \label{fig:mwoz_dsat_feature_dist}
\end{subfigure}
\vspace{-6mm}
\caption{Satisfaction/Dissatisfaction Conversational Pattern Distributions.}
\label{fig:pattern_variance}
\vspace{-3mm}
\end{figure}

Figure~\ref{fig:pattern_variance} reports the satisfaction and dissatisfaction rubric items summarized from the Bing Copilot dataset in the top row, and the bottom row shows the rubric items learned from the MWOZ dataset. Different types of conversational patterns can be observed for the two different conversational systems. 
Each bar indicates the distribution of the number of times that each rubric item appears in a conversation. Because Bing Copilot is a general-purpose conversational system, the summarized rubric items are general conversational patterns. The detailed description of each Bing Copilot rubric item is shown in Table~\ref{tab:copilot_rubric} in Appendix~\ref{app:rubrics}. In contrast, since MWOZ is a booking chatbot, some satisfaction patterns, e.g. booking confirmation or dissatisfaction patterns and plan adaption, are specific to the booking chatbot. The descriptions for each rubric item learned from the MWOZ dataset are listed in Table~\ref{tab:mwoz_rubric} in Appendix~\ref{app:rubrics}.

Similarly, different conversational systems have different service targets, and therefore, the reasons causing user satisfaction or dissatisfaction are related to the target of the system. Because Bing Copilot is a general-purpose question-answering system, inaccurate information contributes to a larger portion of  dissatisfaction. While MWOZ is a booking-reservation system, more of the dissatisfaction is due to a lack of proactivity or a compromise in preference, which means that users have to actively search or choose an option that is less preferred.

\subsection{Knowledge Distillation.} Although \modelname{} can be effectively applied to predict user satisfaction as shown above,
since \modelname{} requires GPT-4 prompting, it is still inefficient to apply USE at web scale (e.g., there have been more than 5 billion conversations in Bing Copilot to date~\cite{convo_nums}. To address this, we propose a knowledge distillation process for each of the rubric items to reduce the cost of the evaluation process. Given the rubric item, we prompt GPT-4 to label a set of conversations for training (the label represents whether or not the conversational pattern described by the rubric item appears in the conversation). Then we calculate an embedding for the conversation (e.g., using OpenAI ada-002) and train a classifier (Logistic Regression) to distill  GPT-4 knowledge (i.e., learn a mapping from embedding to rubric label).

\begin{figure}
\begin{subfigure}{0.5\linewidth}
  \centering
  \includegraphics[clip,width=\linewidth]{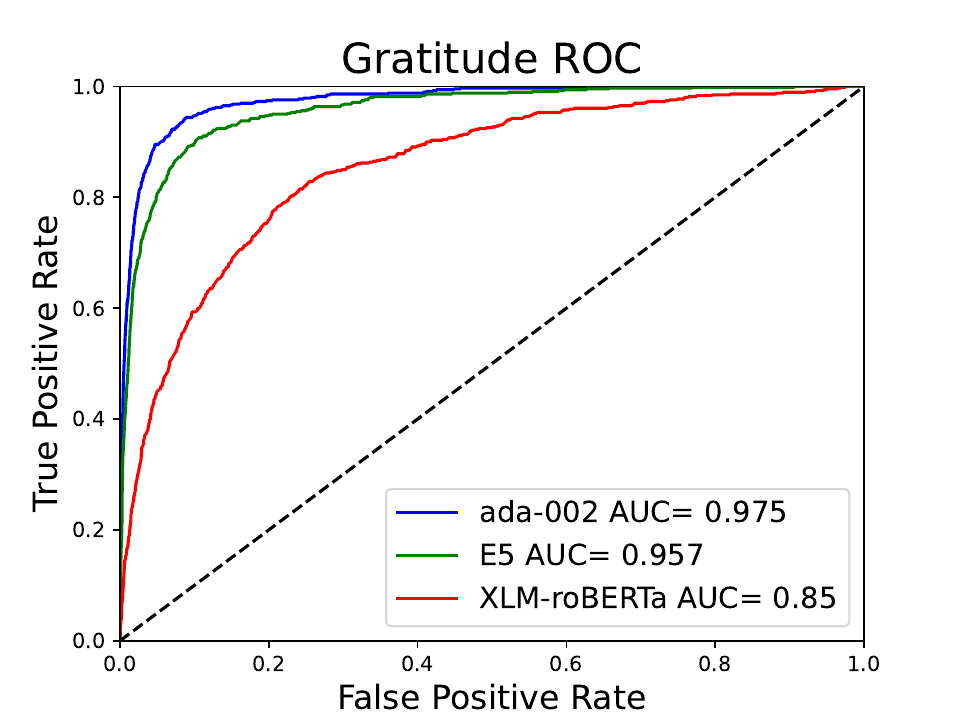}
  \caption{Gratitude.}
  \label{fig:s1_auc}
\end{subfigure}%
\begin{subfigure}{0.5\linewidth}
  \centering
  \includegraphics[clip,width=\linewidth]{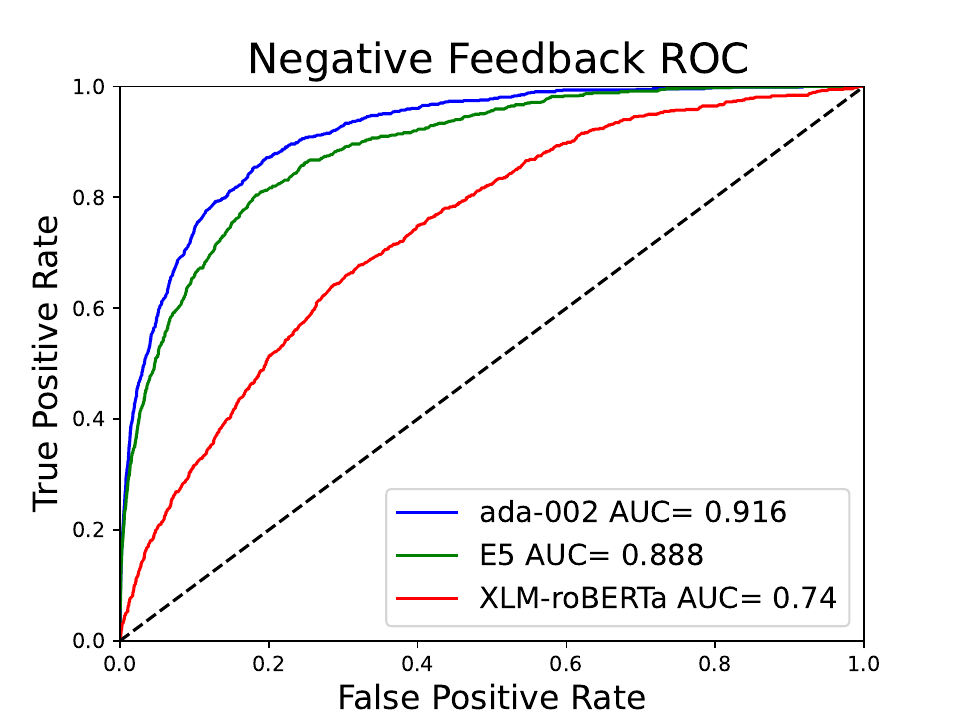}
  \caption{Negative Feedback.}
  \label{fig:d1_auc}
\end{subfigure}%
% \begin{subfigure}{.33\linewidth}
%   \centering
%   \includegraphics[clip,width=\linewidth]{fig/coverage.pdf}
%   \caption{Feedback coverage from Nov 26 to Dec 2 in 2023. \jay{this caption is out of date}}
%   \label{fig:coverage}
% \end{subfigure}
\vspace{-2mm}
\caption{ROC on Knowledge Distillation from GPT-4.}
\label{fig:kd_performance}
\vspace{-2mm}
\end{figure}

We use the above process to distill knowledge from GPT-4 for one of the satisfaction rubric items (Gratitude) and one of the dissatisfaction rubric items (Negative Feedback). Specifically, we train a Gratitude classifier and a Negative-Feedback classifier. The effectiveness of knowledge distillation is shown in Figure~\ref{fig:s1_auc} and Figure~\ref{fig:d1_auc}. A higher AUC metric indicates that the classifier can successfully distill the knowledge from GPT-4 for the given rubric item. We compare the performance of the distilled model with two different embeddings: OpenAI's text-embedding-ada-002~\cite{openai_embeddings} and multilingual E5~\cite{hugging_e5}. As a baseline we compare to an embedding-based sentiment classifier: XLM-roBERTa~\cite{hugging_roberta}.
The results show that ada-002 is the most effective text embedding model for knowledge distillation, so we use that in the experiments below.

\begin{figure}[t]
  \centering
  \includegraphics[clip,width=0.8\linewidth]{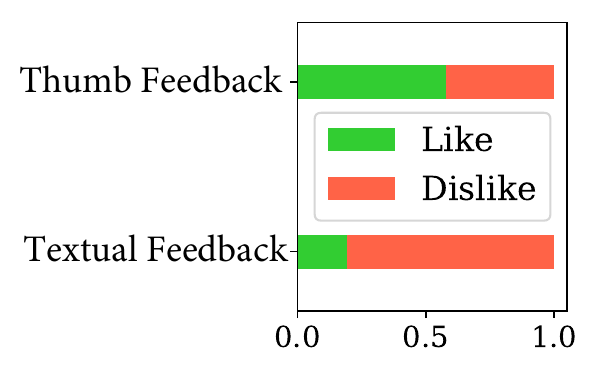}
  \vspace{-4mm}
  \caption{Distributions of Click and Textual Feedback.}
  \vspace{-4mm}
\label{fig:feedback}
\end{figure}

\vspace{2mm}
\noindent\textbf{Feedback Distributions.} After learning the two textual feedback classifiers, we deploy them to a production environment and seek to understand whether they provide different coverage compared to explicit thumb feedback (i.e.,``Like'' or ``Dislike''). Figure~\ref{fig:feedback} reports the distribution of the two types of feedback from one week in production. ``Textual'' feedback records the proportion of conversations that have true labels predicted by the Gratitude classifier (Textual Like) or by the Negative-Feedback classifier (Textual Dislike). Instead of reporting absolute numbers, we report results relative to the proportion of thumb feedback we observe in the data. Figure~\ref{fig:feedback} shows the relative frequency of thumb vs. textual feedback. We can observe that users give more positive feedback through thumb feedback and more negative feedback through their utterances. This also demonstrates the importance of mining conversational SAT/DSAT patterns via \modelname{}.

% \begin{table}[t]
% \caption{Relative comparisons between click feedback (Like or Dislike) and textual feedback (SAT or DSAT)}
% \setlength{\tabcolsep}{1.8pt}
% \begin{center}
% \begin{tabular}{>{\small}l>{\small}c>{\small}c>{\small}c>{\small}c}
% \hline
%  & SAT/Like & DSAT/Dislike & Textual/Both & Click/Both \\
% \hline
% Gains & 0.34 & 1.95 & 20.98 & 20.51  \\
% \hline
% \end{tabular}
% \vspace{-4mm}
% \label{tab:coverage}
% \end{center}
% \end{table}

\subsection{Rubrics as Features}
\begin{figure}[t]
  \centering
  \includegraphics[clip,width=\linewidth]{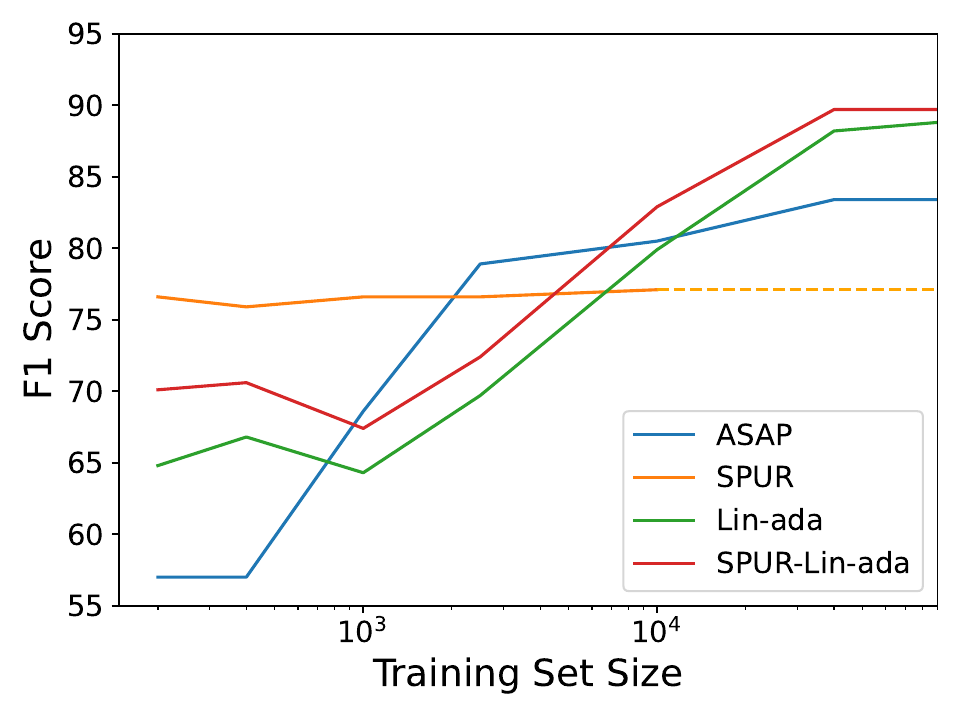}
  \vspace{-1mm}
  \caption{Comparison of F1 scores for the proposed \modelname{} and \modelname{}-Lin-ada models and baseline models for different training set sizes.
  }
  \vspace{-3mm}
\label{fig:model_scale}
\end{figure}

Finally, we seek to understand if combining the rubrics with conversation text embeddings can produce better results using the model proposed in Appendix~\ref{sec:use_model}. We use Bing Copilot dataset with 100K conversations for this experiment. This experiment varies the training size from $400$ to $90K$ of the data and $10K$ of the data is for testing. The results in Figure~\ref{fig:model_scale} indicate that \modelname{} provides the best F1 results for smaller training set sizes. As the training set size increases, the weighted F1 scores of the \modelname{}-Lin-ada (\modelname{} rubrics and linear regression with OpenAI ada-002 embeddings) improves compared to our \modelname{} model and the SOTA embedding ASAP baseline. The results demonstrate that adding the \modelname{} metrics to the feature vector consistently provides additional USE signals that are not captured by the conversation embeddings.  Note, due to the prohibitive cost, we did not retrain \modelname{} for larger training set sized above 10,000 samples. Thus, the orange dashed line from 10K to 90K training samples indicates the \modelname{} F1 score for the test set if we only trained with 10K samples.

\section{Conclusion and Limitations}
In this paper, we proposed \modelfullname{} (\modelname{}), a novel framework for estimating user satisfaction with LLMs in conversational systems. We demonstrated that \modelname{} outperforms existing methods on user satisfaction estimation across different types of conversational systems and also provided insights into the factors that influence user satisfaction. Moreover, \modelname{} is more interpretable because it automatically grounds/scores the dimensions of satisfaction in observed user behavior from Rubric Summarization prompting. We also demonstrated the utility of our rubrics for knowledge distillation and coverage analysis. Finally, we showed the utility of our model for different training set sizes by combining the rubric item scores with the conversational embeddings as features and observed that these rubrics provide extra signals for performance improvement on USE.

\vspace{2mm}
\noindent\textbf{{Limitations}.}
Although \modelname{} outperforms baseline models with limited training sets, an important factor, the framework is costly if the goal is to estimate user satisfaction at the scale of millions of conversations. We have proposed a method to distill knowledge from GPT-4, but a thorough study is needed to show the robustness of this approach. In future work, we will focus on the scalability issues of \modelname{} to reduce its cost at a larger scale.

\newpage
%%
%% The next two lines define the bibliography style to be used, and
%% the bibliography file.
\bibliography{2sample-base}

\begin{thebibliography}{33}
\expandafter\ifx\csname natexlab\endcsname\relax\def\natexlab#1{#1}\fi

\bibitem[{Ada()}]{openai_embeddings}
OpenAI Ada.
\newblock \href {https://platform.openai.com/docs/guides/embeddings/what-are-embeddings} {Embeddings}.
\newblock Accessed on: Feb 15, 2024.

\bibitem[{Bai et~al.(2022)Bai, Kadavath, Kundu, Askell, Kernion, Jones, Chen, Goldie, Mirhoseini, McKinnon, Chen, Olsson, Olah, Hernandez, Drain, Ganguli, Li, Tran{-}Johnson, Perez, Kerr, Mueller, Ladish, Landau, Ndousse, Lukosiute, Lovitt, Sellitto, Elhage, Schiefer, Mercado, DasSarma, Lasenby, Larson, Ringer, Johnston, Kravec, Showk, Fort, Lanham, Telleen{-}Lawton, Conerly, Henighan, Hume, Bowman, Hatfield{-}Dodds, Mann, Amodei, Joseph, McCandlish, Brown, and Kaplan}]{rlaif}
Yuntao Bai, Saurav Kadavath, Sandipan Kundu, Amanda Askell, Jackson Kernion, Andy Jones, Anna Chen, Anna Goldie, Azalia Mirhoseini, Cameron McKinnon, Carol Chen, Catherine Olsson, Christopher Olah, Danny Hernandez, Dawn Drain, Deep Ganguli, Dustin Li, Eli Tran{-}Johnson, Ethan Perez, Jamie Kerr, Jared Mueller, Jeffrey Ladish, Joshua Landau, Kamal Ndousse, Kamile Lukosiute, Liane Lovitt, Michael Sellitto, Nelson Elhage, Nicholas Schiefer, Noem{\'{\i}} Mercado, Nova DasSarma, Robert Lasenby, Robin Larson, Sam Ringer, Scott Johnston, Shauna Kravec, Sheer~El Showk, Stanislav Fort, Tamera Lanham, Timothy Telleen{-}Lawton, Tom Conerly, Tom Henighan, Tristan Hume, Samuel~R. Bowman, Zac Hatfield{-}Dodds, Ben Mann, Dario Amodei, Nicholas Joseph, Sam McCandlish, Tom Brown, and Jared Kaplan. 2022.
\newblock Constitutional {AI:} harmlessness from {AI} feedback.
\newblock \emph{CoRR}, abs/2212.08073.

\bibitem[{Bodigutla et~al.(2019)Bodigutla, Polymenakos, and Matsoukas}]{rq_use}
Praveen~Kumar Bodigutla, Lazaros Polymenakos, and Spyros Matsoukas. 2019.
\newblock Multi-domain conversation quality evaluation via user satisfaction estimation.
\newblock \emph{CoRR}, abs/1911.08567.

\bibitem[{Bodigutla et~al.(2020)Bodigutla, Tiwari, Valls{-}Vargas, Polymenakos, and Matsoukas}]{rq_use2}
Praveen~Kumar Bodigutla, Aditya Tiwari, Josep Valls{-}Vargas, Lazaros Polymenakos, and Spyros Matsoukas. 2020.
\newblock Joint turn and dialogue level user satisfaction estimation on multi-domain conversations.
\newblock \emph{CoRR}, abs/2010.02495.

\bibitem[{Cai and Chen(2020)}]{use_rec}
Wanling Cai and Li~Chen. 2020.
\newblock Predicting user intents and satisfaction with dialogue-based conversational recommendations.
\newblock In \emph{Proceedings of the 28th {ACM} Conference on User Modeling, Adaptation and Personalization, {UMAP} 2020, Genoa, Italy, July 12-18, 2020}, pages 33--42. {ACM}.

\bibitem[{Christiano et~al.(2017)Christiano, Leike, Brown, Martic, Legg, and Amodei}]{rlhf}
Paul~F. Christiano, Jan Leike, Tom~B. Brown, Miljan Martic, Shane Legg, and Dario Amodei. 2017.
\newblock Deep reinforcement learning from human preferences.
\newblock In \emph{Advances in Neural Information Processing Systems 30: Annual Conference on Neural Information Processing Systems}, pages 4299--4307.

\bibitem[{Deng et~al.(2022)Deng, Zhang, Lam, Cheng, and Meng}]{usda}
Yang Deng, Wenxuan Zhang, Wai Lam, Hong Cheng, and Helen Meng. 2022.
\newblock \href {https://doi.org/10.1145/3485447.3512020} {User satisfaction estimation with sequential dialogue act modeling in goal-oriented conversational systems}.
\newblock In \emph{{WWW} '22: The {ACM} Web Conference 2022, Virtual Event, Lyon, France, April 25 - 29, 2022}, pages 2998--3008. {ACM}.

\bibitem[{Deriu et~al.(2021)Deriu, Rodrigo, Otegi, Echegoyen, Rosset, Agirre, and Cieliebak}]{survey_dialogue_sys}
Jan Deriu, {\'{A}}lvaro Rodrigo, Arantxa Otegi, Guillermo Echegoyen, Sophie Rosset, Eneko Agirre, and Mark Cieliebak. 2021.
\newblock Survey on evaluation methods for dialogue systems.
\newblock \emph{Artif. Intell. Rev.}, 54(1):755--810.

\bibitem[{E5()}]{hugging_e5}
Hugginface E5.
\newblock \href {https://huggingface.co/intfloat/multilingual-e5-large} {intfloat/multilingual-e5-large}.
\newblock Accessed on: Accessed on: Feb 15, 2024.

\bibitem[{Eric et~al.(2020)Eric, Goel, Paul, Sethi, Agarwal, Gao, Kumar, Goyal, Ku, and Hakkani{-}T{\"{u}}r}]{mwoz}
Mihail Eric, Rahul Goel, Shachi Paul, Abhishek Sethi, Sanchit Agarwal, Shuyang Gao, Adarsh Kumar, Anuj~Kumar Goyal, Peter Ku, and Dilek Hakkani{-}T{\"{u}}r. 2020.
\newblock Multiwoz 2.1: {A} consolidated multi-domain dialogue dataset with state corrections and state tracking baselines.
\newblock In \emph{Proceedings of The 12th Language Resources and Evaluation Conference, {LREC} 2020, Marseille, France, May 11-16, 2020}, pages 422--428. European Language Resources Association.

\bibitem[{Hu et~al.(2023)Hu, Feng, Luu, Hooi, and Lipani}]{urgo}
Zhiyuan Hu, Yue Feng, Anh~Tuan Luu, Bryan Hooi, and Aldo Lipani. 2023.
\newblock Unlocking the potential of user feedback: Leveraging large language model as user simulators to enhance dialogue system.
\newblock In \emph{Proceedings of the 32nd {ACM} International Conference on Information and Knowledge Management, {CIKM} 2023, Birmingham, United Kingdom, October 21-25, 2023}, pages 3953--3957. {ACM}.

\bibitem[{Huang et~al.(2023)Huang, Mamidanna, Jangam, Zhou, and Gilpin}]{faith_explanation}
Shiyuan Huang, Siddarth Mamidanna, Shreedhar Jangam, Yilun Zhou, and Leilani~H. Gilpin. 2023.
\newblock \href {https://doi.org/10.48550/ARXIV.2310.11207} {Can large language models explain themselves? {A} study of llm-generated self-explanations}.
\newblock \emph{CoRR}, abs/2310.11207.

\bibitem[{Kachuee et~al.(2021{\natexlab{a}})Kachuee, Yuan, Kim, and Lee}]{use_cl}
Mohammad Kachuee, Hao Yuan, Young{-}Bum Kim, and Sungjin Lee. 2021{\natexlab{a}}.
\newblock Self-supervised contrastive learning for efficient user satisfaction prediction in conversational agents.
\newblock In \emph{Proceedings of the 2021 Conference of the North American Chapter of the Association for Computational Linguistics: Human Language Technologies, {NAACL-HLT} 2021, Online, June 6-11, 2021}, pages 4053--4064. Association for Computational Linguistics.

\bibitem[{Kachuee et~al.(2021{\natexlab{b}})Kachuee, Yuan, Kim, and Lee}]{use_2}
Mohammad Kachuee, Hao Yuan, Young-Bum Kim, and Sungjin Lee. 2021{\natexlab{b}}.
\newblock Self-supervised contrastive learning for efficient user satisfaction prediction in conversational agents.
\newblock In \emph{Proceedings of the 2021 Conference of the North American Chapter of the Association for Computational Linguistics: Human Language Technologies}, Online. Association for Computational Linguistics.

\bibitem[{Kirk et~al.(2023)Kirk, Vidgen, R{\"{o}}ttger, and Hale}]{rlhf_risk}
Hannah~Rose Kirk, Bertie Vidgen, Paul R{\"{o}}ttger, and Scott~A. Hale. 2023.
\newblock Personalisation within bounds: {A} risk taxonomy and policy framework for the alignment of large language models with personalised feedback.
\newblock \emph{CoRR}, abs/2303.05453.

\bibitem[{Kojima et~al.(2022)Kojima, Gu, Reid, Matsuo, and Iwasawa}]{zero_shot_reasoner}
Takeshi Kojima, Shixiang~Shane Gu, Machel Reid, Yutaka Matsuo, and Yusuke Iwasawa. 2022.
\newblock Large language models are zero-shot reasoners.
\newblock In \emph{NeurIPS}.

\bibitem[{Lampinen et~al.(2022)Lampinen, Dasgupta, Chan, Mathewson, Tessler, Creswell, McClelland, Wang, and Hill}]{fewshot}
Andrew~K. Lampinen, Ishita Dasgupta, Stephanie C.~Y. Chan, Kory~W. Mathewson, Michael~Henry Tessler, Antonia Creswell, James~L. McClelland, Jane Wang, and Felix Hill. 2022.
\newblock Can language models learn from explanations in context?
\newblock In \emph{Findings of the Association for Computational Linguistics: {EMNLP} 2022, Abu Dhabi, United Arab Emirates, December 7-11, 2022}, pages 537--563. Association for Computational Linguistics.

\bibitem[{Liang et~al.(2021)Liang, Takanobu, Li, Zhang, Chen, and Huang}]{use_1}
Runze Liang, Ryuichi Takanobu, Feng-Lin Li, Ji~Zhang, Haiqing Chen, and Minlie Huang. 2021.
\newblock Turn-level user satisfaction estimation in {E}-commerce customer service.
\newblock In \emph{Proceedings of the 4th Workshop on e-Commerce and NLP}, pages 26--32, Online. Association for Computational Linguistics.

\bibitem[{Mehdi()}]{convo_nums}
Yusuf Mehdi.
\newblock \href {https://blogs.microsoft.com/blog/2024/01/15/bringing-the-full-power-of-copilot-to-more-people-and-businesses/} {Bringing the full power of copilot to more people and businesses}.
\newblock Accessed on: Feb 15, 2024.

\bibitem[{Pan et~al.(2022)Pan, Ma, Pflugfelder, and Groh}]{use_3}
Yan Pan, Mingyang Ma, Bernhard Pflugfelder, and Georg Groh. 2022.
\newblock User satisfaction modeling with domain adaptation in task-oriented dialogue systems.
\newblock In \emph{Proceedings of the 23rd Annual Meeting of the Special Interest Group on Discourse and Dialogue}, Edinburgh, UK. Association for Computational Linguistics.

\bibitem[{Rastogi et~al.(2020)Rastogi, Zang, Sunkara, Gupta, and Khaitan}]{sgd}
Abhinav Rastogi, Xiaoxue Zang, Srinivas Sunkara, Raghav Gupta, and Pranav Khaitan. 2020.
\newblock \href {https://doi.org/10.1609/AAAI.V34I05.6394} {Towards scalable multi-domain conversational agents: The schema-guided dialogue dataset}.
\newblock In \emph{The Thirty-Fourth {AAAI} Conference on Artificial Intelligence, {AAAI} 2020, The Thirty-Second Innovative Applications of Artificial Intelligence Conference, {IAAI} 2020, The Tenth {AAAI} Symposium on Educational Advances in Artificial Intelligence, {EAAI} 2020, New York, NY, USA, February 7-12, 2020}, pages 8689--8696. {AAAI} Press.

\bibitem[{reward deberta()}]{hugging_reward}
Hugginface reward deberta.
\newblock \href {https://huggingface.co/OpenAssistant/reward-model-deberta-v3-large-v2} {Openassistant/reward-model-deberta-v3-large-v2}.
\newblock Accessed on: Accessed on: Feb 15, 2024.

\bibitem[{Schmitt and Ultes(2015)}]{iq}
Alexander Schmitt and Stefan Ultes. 2015.
\newblock Interaction quality: Assessing the quality of ongoing spoken dialog interaction by experts - and how it relates to user satisfaction.
\newblock \emph{Speech Commun.}, 74:12--36.

\bibitem[{Siro et~al.(2022)Siro, Aliannejadi, and de~Rijke}]{redial}
Clemencia Siro, Mohammad Aliannejadi, and Maarten de~Rijke. 2022.
\newblock Understanding user satisfaction with task-oriented dialogue systems.
\newblock In \emph{{SIGIR} '22: The 45th International {ACM} {SIGIR} Conference on Research and Development in Information Retrieval, Madrid, Spain, July 11 - 15, 2022}, pages 2018--2023. {ACM}.

\bibitem[{Song et~al.(2019)Song, Bing, Gao, Lin, Zhao, Wang, Sun, Liu, and Zhang}]{sentiment}
Kaisong Song, Lidong Bing, Wei Gao, Jun Lin, Lujun Zhao, Jiancheng Wang, Changlong Sun, Xiaozhong Liu, and Qi~Zhang. 2019.
\newblock Using customer service dialogues for satisfaction analysis with context-assisted multiple instance learning.
\newblock In \emph{Proceedings of the 2019 Conference on Empirical Methods in Natural Language Processing and the 9th International Joint Conference on Natural Language Processing, {EMNLP-IJCNLP}}, pages 198--207. Association for Computational Linguistics.

\bibitem[{Song et~al.(2023)Song, Kang, Liu, Li, Sun, and Liu}]{use_speaker}
Kaisong Song, Yangyang Kang, Jiawei Liu, Xurui Li, Changlong Sun, and Xiaozhong Liu. 2023.
\newblock A speaker turn-aware multi-task adversarial network for joint user satisfaction estimation and sentiment analysis.
\newblock In \emph{Thirty-Seventh {AAAI} Conference on Artificial Intelligence, {AAAI} 2023, Thirty-Fifth Conference on Innovative Applications of Artificial Intelligence, {IAAI} 2023, Thirteenth Symposium on Educational Advances in Artificial Intelligence, {EAAI} 2023, Washington, DC, USA, February 7-14, 2023}, pages 13582--13590. {AAAI} Press.

\bibitem[{Sun et~al.(2021)Sun, Zhang, Balog, Ren, Ren, Chen, and de~Rijke}]{uss}
Weiwei Sun, Shuo Zhang, Krisztian Balog, Zhaochun Ren, Pengjie Ren, Zhumin Chen, and Maarten de~Rijke. 2021.
\newblock Simulating user satisfaction for the evaluation of task-oriented dialogue systems.
\newblock In \emph{{SIGIR} '21: The 44th International {ACM} {SIGIR} Conference on Research and Development in Information Retrieval, Virtual Event, Canada, July 11-15, 2021}, pages 2499--2506. {ACM}.

\bibitem[{Sun et~al.(2023)Sun, Li, Li, Wu, Guo, Zhang, and Wang}]{carp}
Xiaofei Sun, Xiaoya Li, Jiwei Li, Fei Wu, Shangwei Guo, Tianwei Zhang, and Guoyin Wang. 2023.
\newblock Text classification via large language models.
\newblock In \emph{Findings of the Association for Computational Linguistics: {EMNLP} 2023, Singapore, December 6-10, 2023}, pages 8990--9005. Association for Computational Linguistics.

\bibitem[{Walker et~al.(1997)Walker, Litman, Kamm, and Abella}]{paradise}
Marilyn~A. Walker, Diane~J. Litman, Candace~A. Kamm, and Alicia Abella. 1997.
\newblock {PARADISE:} {A} framework for evaluating spoken dialogue agents.
\newblock In \emph{35th Annual Meeting of the Association for Computational Linguistics and 8th Conference of the European Chapter of the Association for Computational Linguistics, Proceedings of the Conference, 7-12 July 1997, Universidad Nacional de Educaci{\'{o}}n a Distancia (UNED), Madrid, Spain}, pages 271--280. Morgan Kaufmann Publishers / {ACL}.

\bibitem[{Wei et~al.(2022)Wei, Wang, Schuurmans, Bosma, Ichter, Xia, Chi, Le, and Zhou}]{cot}
Jason Wei, Xuezhi Wang, Dale Schuurmans, Maarten Bosma, Brian Ichter, Fei Xia, Ed~H. Chi, Quoc~V. Le, and Denny Zhou. 2022.
\newblock Chain-of-thought prompting elicits reasoning in large language models.
\newblock In \emph{Advances in Neural Information Processing Systems 35: Annual Conference on Neural Information Processing Systems 2022, NeurIPS}.

\bibitem[{XLM-roBERTa()}]{hugging_roberta}
Hugginface XLM-roBERTa.
\newblock \href {https://huggingface.co/cardiffnlp/twitter-xlm-roberta-base-sentiment} {cardiffnlp/twitter-xlm-roberta-base-sentiment}.
\newblock Accessed on: Accessed on: Feb 15, 2024.

\bibitem[{Ye and Durrett(2023)}]{opt_explanation}
Xi~Ye and Greg Durrett. 2023.
\newblock Explanation selection using unlabeled data for chain-of-thought prompting.
\newblock In \emph{Proceedings of the 2023 Conference on Empirical Methods in Natural Language Processing, {EMNLP} 2023, Singapore, December 6-10, 2023}, pages 619--637. Association for Computational Linguistics.

\bibitem[{Ye~Fanghua(2023)}]{asap}
Yilmaz~Emine Ye~Fanghua, Hu~Zhiyuan. 2023.
\newblock Modeling user satisfaction dynamics in dialogue via hawkes process.
\newblock In \emph{The 61st Annual Meeting of the Association for Computational Linguistics (ACL’23)}.

\end{thebibliography}

%%
%% If your work has an appendix, this is the place to put it.
\appendix

\section{Prompts}

\subsection{Supervised Extraction Prompt}
\label{apx:se_prompt}
\begin{verbatim}
You job is to understand and elaborate how a 
user expresses that they are **satisfied** 
with their interaction with an AI agent. You 
will be given a conversation that a user had 
with an AI agent where the user provided a 
signal of satisfaction through a like button.

Your task is to summarize how the user expressed 
satisfaction with the conversation. 
Instructions:
- Provide your answer in xml format between 
<REASONS></REASONS> tags.
- Return NONE if you can't think of any part 
of the user's utterances that expresses 
satisfaction.
- The reasons you summarized should be 
grounded  on the conversation history only. 
You should **NOT** extrapolate, imagine, or 
hallucinate  beyond the text of the 
conversation that is given.
- The reasons should be mutually exclusive.
- You should **NOT** refer to the fact that 
there was a like in your summary.
- Your summary should be concise, use bullet 
points, and provide no more than 3 reasons.

<CONVERSATION>
[user-agent utterances]
</CONVERSATION>

The main reasons why the user is satisfied 
with the interaction are:
\end{verbatim}

\subsection{Rubric Summarization Prompt}
\label{apx:sr_prompt}
\begin{verbatim}
# Task 
You job is to summarize why a user feels 
**satisfied** with their interaction with 
an AI agent and provide a rubric for 
evaluation of a single conversation. You 
will be given a list of example explanations 
from conversations that users had with an 
AI agent where these users provided a 
signal of satisfaction.

# Instruction 
Your task is to provide a rubric to 
identify user satisfaction with respect 
to a conversation. Requirements:
* Provide your answer as a numbered list 
of up to {num_rubric} bullet items.
* The rubric should be user-centric, 
concise, and mutually exclusive.


# Example Explanations of User Satisfaction
"[S_b + n-item rubrics from S_{b-1}. 
If b=0, put S_0]"


# Now summarize these examples into a 
rubric to identify user satisfaction with 
respect to a conversation. Requirements:
* Provide your answer as a numbered list 
of up to {num_rubric} bullet items.
* The number of items in the rubric should 
be less than {num_rubric}.
* The rubric should be user-centric, 
concise, and mutually exclusive.
* Provide your answer as a numbered list of 
bullet items in <Rubric></Rubric>. The 
output format is as follows:
```
# Output
<Rubric>
1. [item 1]
2. [item 2]
3. [item 3]
...
</Rubric>
```

# Output
\end{verbatim}

\subsection{User Satisfaction Estimation Prompt}
\label{apx:use_prompt}
\begin{verbatim}
# Your task is to evaluate both user 
satisfaction and dissatisfaction with a 
conversational AI agent by applying the 
given rubrics to the given conversation 
history between the user and the agent.

# Rubric instructions
- Each rubric contains 10 criteria.
- Each criterion has a Yes or No statement.
- Your job is to go through the 
conversation history carefully and answer 
Y to each statement that applies to the 
user utterances in the conversation, then 
give the statement a score of 1-10 to 
reflect how likely the expressed sentiment 
will impact the user's overall 
satisfaction/dissatisfaction with the 
interaction. If the statement is not 
applicable answer N and give an overall 
score of 0.
- Each rubric is formatted in a table format
with 10 rows and two columns: Index|Y/N 
Question.

# SATISFACTION RUBRIC
{n_item_sat_rubric}

# DISSATISFACTION RUBRIC
{n_item_dsat_rubric}

# Task:
- Go through the conversation history 
thoroughly and evaluate the user's 
utterances. Do not consider the AI's 
responses except to put the user's 
response in context.
- For each rubric question think about your 
answer to each question carefully.
- Answer Y or N only to each rubric question.
- For Y answer, score your answer on a scale 
of 1-10 (low to high) to reflect how likely 
the expressed sentiment will impact the 
user's overall satisfaction or 
dissatisfaction with the interaction. 
For N answer, score 0.
- Only provide ONE most confident answer to 
each question.
- You *MUST* output your answers to all 10 
questions provided in each rubric.

# Conversation:
[user-agent utternaces]

# Answers
\end{verbatim}

\section{Labeling Adjustment for the Open Data}
\label{apx:preprocess}

The open datasets include turn-by-turn labels whereas \modelname{} requires a label for the entire conversation.
The process of translating turn-by-turn labels into conversation labels follows these steps:
\begin{itemize}
    \item If the full conversation has only neutral and SAT, then the label for full conversation is SAT.
    \item If the full conversation has only neutral and DSAT, then the label for full conversation is DSAT.
    \item If the full conversation has only neutral, then the label for the full conversation is neutral.
    \item If the full conversation has both SAT and DSAT.
    \begin{itemize}
        \item start from the beginning of the conversation, discard the rest of the conversation when contradiction happens 
        and assign the label as the first non-neutral label. 
    \end{itemize}
\end{itemize}

The modified label counts for the three open datasets after following this label conversion process are provided in Table~\ref{tab:label_dist}.
\begin{table}[htbp]
\caption{Label Distribution}
\vspace{-4mm}
\begin{center}
\begin{tabular}{lllll}
\hline
Dataset  & SAT & DSAT & Neutral & Sum \\
\hline
redial & 822 & 463 & 102 & 1387 \\
sgd & 1008 & 496 & 179 & 1683 \\
mwoz & 560 & 524 & 71 & 1155  \\
\hline
\end{tabular}
\label{tab:label_dist}
\end{center}
\end{table}

\section{Experiment Setup}
We use GPT-4 for the entire process of training and evaluating \modelname{}, and \modelname{}-Lin-ada is trained and tested on an NVIDIA A100 instance. Every experiment runs one time but with a large testing size (80\% is used for testing). The hyperparameters are listed as follows:
\begin{itemize}
    \item The number of top-k SAT or DSAT patterns for a conversation is $3$.
    \item The batch size for each minibatch is $100$ SAT/DSAT patterns.
    \item The number of items for the satisfaction rubric and dissatisfaction rubric is $10$.
\end{itemize}

\section{User Satisfaction Model}
\label{sec:use_model}
The User Satisfaction Rubrics can be used by themselves to compute a USE score. However, we have found that the utility can be further improved by including a text embedding of the chat conversation in addition to the values of the rubrics. In particular, results show that using the OpenAI ada-002 text embeddings are particularly effective.

\begin{figure}[tbh]
  \centering
  \includegraphics[trim=2.5in 2.0in 0.5in 1.0in, clip, width=3.75in]{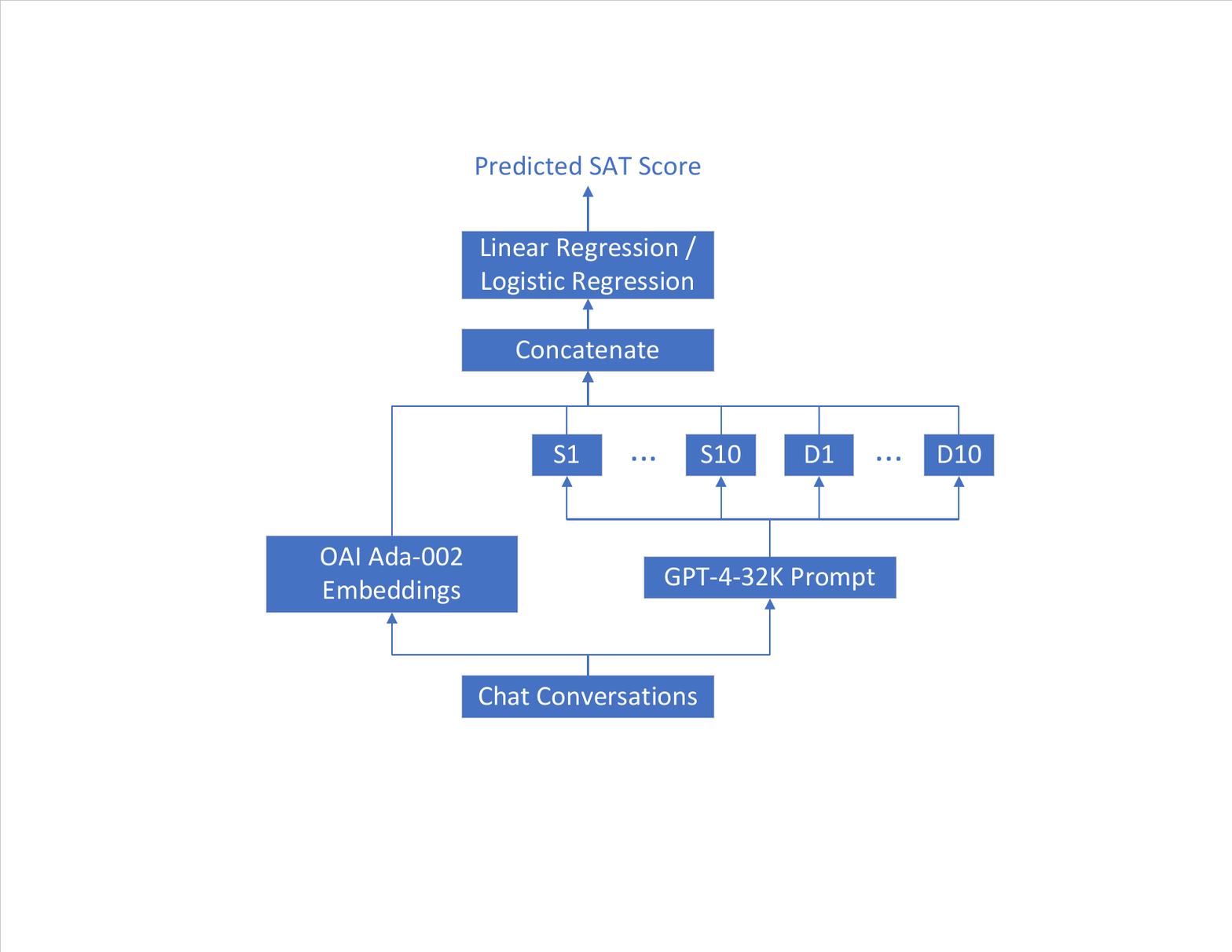}
  \caption{The proposed model combines the \modelname{} LLM rubrics and conversation embeddings.}
\label{fig:combllmrub_convemb}
\end{figure}

The proposed model is depicted in Figure~\ref{fig:combllmrub_convemb}. On the left, the conversations are projected into an embedding space using the GPT-3 Ada-002 embeddings. In parallel, the 20 LLM rubric itmes are computed using the GPT-4-32K LLM on the right. The 1536-dimension conversation embedding vector is concatenated with the 20 \modelname{} rubric scores to form the final feature vector which is then input to a model such as Linear Regression, Logistic Regression, or a DNN. The output of the model is the final predicted USE score.

\begin{figure}[tbh]
  \centering
  \includegraphics[clip,width=\linewidth]{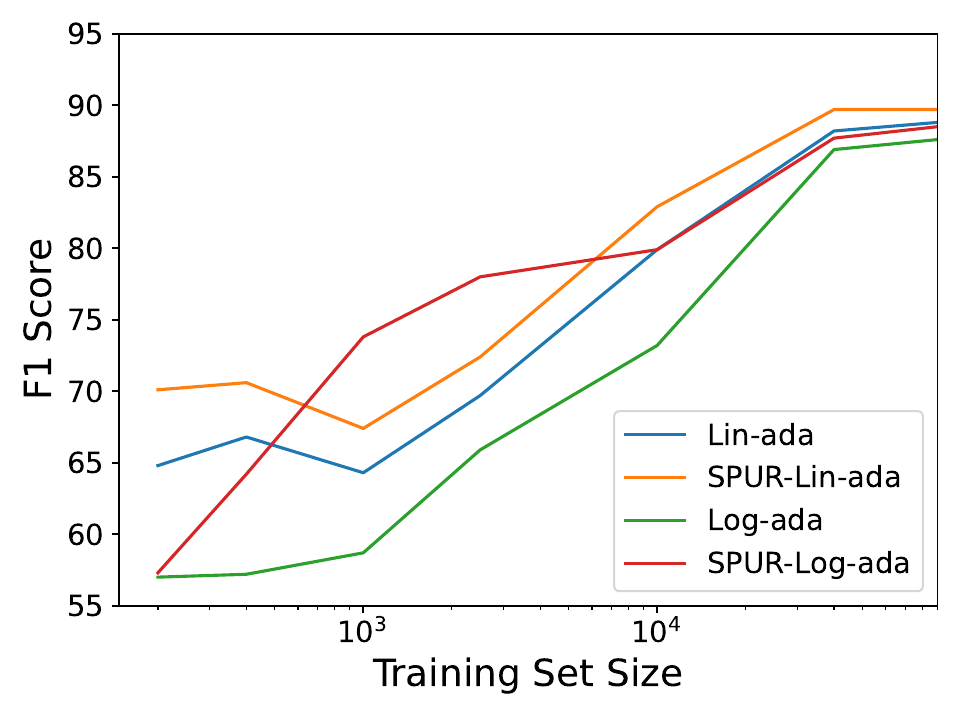}
  \caption{Comparison of F1 scores for the proposed \modelname{} and the combined \modelname{} and conversation embedding models for different training set sizes.
  Using logistic regression offers better performance for smaller training set sizes, but linear regression yields the best results for the higher range.}
\label{fig:model_scale_2}
\end{figure}

Figure~\ref{fig:model_scale_2} compares the results using a final linear regression layer and a logistic regression layer, with and without the \modelname{} rubrics. The figure shows that adding the \modelname{} rubrics improves both baseline models which only consider the conversation embeddings as features. Furthermore, while the two logistic regression models offer the best performance for smaller training set sized, the linear regression models are the best performing models for the larger training set sizes. We also evaluated replacing the regression layer (e.g., linear, logistic) with a DNN, but the performance was much worse due to overfitting.

\section{Usage of AI Assistants}
\label{app:rubrics}
\modelname{} is an implementation based on GPT-4. We only use Bing Copilot to assist our writing to identify grammar errors, typos and rephrase terms for readability.
\begin{table*}[htbp]
\caption{Satisfaction and Dissatisfaction Features for Copilot}
\vspace{-4mm}
\begin{center}
\begin{tabular}{p{0.15\linewidth}p{0.3\linewidth}|p{0.15\linewidth}p{0.3\linewidth}}
\hline
 \multicolumn{2}{c}{Satisfaction} & \multicolumn{2}{c}{Dissatisfaction}\\ \cmidrule(l{0.5em}r{0.5em}){1-2}\cmidrule(l{0.5em}r{0.5em}){3-4}
 Name & Description &Name & Description\\
\hline
Gratitude & The user thanks or compliments the AI agent for its help, quality, performance, or abilities. & Repetition & The user repeats their query or request multiple times. \\
Positive Feedback & The user expresses positive emotions or evaluations using words, phrases, punctuation marks, or emoticons. & Errors & The user points out an error, inconsistency, or inaccuracy in the AI's output or information and does not receive any acknowledgment or apology from the agent.\\
Engagement & The user engages in a diverse and lengthy conversation with the AI agent, covering multiple topics or domains. & Negative Feedback &  The user uses a negative tone or words to express frustration, disappointment, anger, or disrespect towards the AI agent.  \\
Follow-up & The user asks follow-up questions or requests more information from the AI agent that show curiosity and interest in learning more. & Topic Switch & The user changes their topic or query abruptly. \\
No Frustration & The user does not express any negative emotion toward the AI agent's responses throughout the conversation. & Lack Visualization &  The user does not receive any visual output from the AI agent when they expect images, links, charts, etc. \\
Suggestion Acceptance & The user accepts or follows the AI agent's suggestions, recommendations, and feedback without hesitation, resistance, or challenging it. & No Engagement & The user does not engage with the AI agent's questions, comments, suggestions, feedback requests, etc. \\
Personal Details & The user initiates or continues a personal conversation with the AI agent by sharing details about themselves or asking how it is doing. & Irrelevant Information & The user receives a generic, vague, irrelevant answer from the AI agent that does not address their specific needs, goals, or preferences. \\
Task Request & The user requests specific tasks from the AI agent that match its domain and scope of knowledge, abilities, skills, and expertise. & Complex Answer & The user receives a long and complex answer from the AI agent that may be overwhelming, confusing, or too technical for them.  \\
Correction & The user corrects some of the AI agent's mistakes, guesses, errors, or misunderstandings in a cooperative, trusting, respectful, and polite manner. & Sudden End & The conversation ends abruptly without fulfilling, completing, or addressing the initial request, problem, task, or goal.  \\
Learning & The user enjoys, appreciates, and learns from different formats, styles, modes, and media of outputs and services, as well as information provided, explained, and generated by the AI agent. & Lack Diversity & The user expects a more interactive, engaging, personalized, humorous, and creative response from the AI, rather than a generic, pre-written, factual, technical, verbose one. \\
\hline
\end{tabular}
\label{tab:copilot_rubric}
\end{center}
\end{table*}

\begin{table*}[htbp]
\caption{Satisfaction and Dissatisfaction Features for MWOZ}
\vspace{-4mm}
\begin{center}
\begin{tabular}{p{0.15\linewidth}p{0.3\linewidth}|p{0.15\linewidth}p{0.3\linewidth}}
\hline
 \multicolumn{2}{c}{Satisfaction} & \multicolumn{2}{c}{Dissatisfaction}\\ \cmidrule(l{0.5em}r{0.5em}){1-2}\cmidrule(l{0.5em}r{0.5em}){3-4}
 Name & Description &Name & Description\\
\hline
Gratitude & The user thanks the AI agent for its service, indicating gratitude and appreciation. & Repetition & The user repeats their query or request multiple times. \\
Booking Acceptance & The user accepts the AI agent's suggestions or bookings without asking for changes or alternatives, implying trust and satisfaction. & Lack Feedback & The user does not receive any confirmation or feedback from the AI after making requests, asking questions, or providing information, leading to uncertainty and confusion.\\
No Frustration & The user does not express any frustration, confusion, or dissatisfaction with the AI agent's responses or queries throughout the conversation. & Irrelevant Information & The user receives irrelevant or incomplete information from the AI that does not align with their queries or expectations, which shows a lack of understanding or flexibility. \\
Follow-up & The user asks questions about the information or options provided by the AI agent, showing interest and engagement. & Ignored or Misunderstood & The user feels ignored or misunderstood by the AI as it does not answer some of their questions, acknowledge their inputs, or provide any clarification. \\
Booking Confirmation & The user confirms their booking details or information with a positive expression, showing agreement and happiness.& Compromise Preference &  The user has to compromise on their desired options or criteria because of limited availability or mismatched recommendations from the AI.\\
Farewell & The user ends the conversation with a polite farewell and no complaints or requests for further assistance. & Lack of Proactivity &The user has to ask basic questions about features or details that the AI should have provided upfront. \\
Cooperation & The user follows the AI agent's guidance and prompts without hesitation or objection, indicating acceptance and cooperation. & Plan Adaption & The user changes their mind about something they previously requested or agreed upon (e.g., location preference) without giving a clear reason. \\
Clarification &  The user specifies their preferences or constraints clearly and specifically, showing confidence and comfort in communicating with the AI agent. & Topic Switch & The user switches to a different topic without closing the previous one.\\
Request Fulfillment &  The user receives relevant and helpful information from the AI agent that matches their requests, such as phone number, price, etc. & Failed bookings & The user experienced several failed bookings and received inconsistent information from the AI about availability.  \\
Request Flexibility & The user is able to change their query or ask for different types of information without encountering any errors or misunderstandings from the AI agent. & False Confirmation& The user was misled by the AI's confirmation messages, which turned out to be false. \\
\hline
\end{tabular}
\label{tab:mwoz_rubric}
\end{center}
\end{table*}

\end{document}